\documentclass[aps,prl,,twocolumn,groupedaddress]{revtex4}
\usepackage{graphicx}
\begin{document}
\title{Fidelity of Quantum Interferometers}
\author{Thomas B. Bahder}
\email{bahder@arl.army.mil}
\author{Paul A. Lopata}
\affiliation{
U.\ S.\ Army Research Laboratory \\
2800 Powder Mill Road \\
Adelphi, Maryland, USA 20783-1197}


\date{February 14, 2006}

\begin{abstract}
For a generic interferometer, the conditional
probability density distribution, $p(\phi|m)$, for the phase $\phi$ given measurement
outcome $m$,  will generally have multiple peaks. Therefore,
the phase sensitivity of an interferometer cannot be adequately characterized by the
standard deviation, such as  $\Delta\phi\sim 1/\sqrt{N}$ (the standard limit),
or  $\Delta\phi\sim 1/N$ (the Heisenberg limit). We propose an alternative measure of phase
sensitivity--the fidelity of an interferometer--defined as the Shannon mutual
information  between the phase shift $\phi$\ and the measurement
outcomes $m$. As an example application of interferometer fidelity, we consider a 
generic optical Mach-Zehnder
interferometer, used as a sensor of a classical field. We find the surprising 
result that an entangled {\it N00N} state input
leads to a lower fidelity than a Fock state input, for the same photon number.

\end{abstract}

\maketitle

{\it Introduction.} Phase sensitivity of interferometers has been a topic of research for many
years because of interest in the fundamental limitations of
measurement~\mbox{\cite{Godun2001,Giovannetti2006}}, gravitational-wave
detection\cite{Thorne1980}, and optical~\cite{Chow1985,Bertocchi2006}, atom~\cite{Zimmer2004}, and 
Bose-Einstein condensate(BEC)-based gyroscopes\cite{Gupta2005,Wang2005,Tolstikhin2005}. 
Recently,
applications to sensors are being explored\cite{Didomenico2004,Kapale2005}.
\ \ The phase sensitivity of interferometers is believed to be limited by
quantum fluctuations\cite{Caves1981}, and the phase sensitivity of various
interferometers has been explored for different types of input states, such
as squeezed states\cite{Caves1981,Bondurant1984}, and  number
states\cite{YurkePRL1986,YurkePRA1986,Yuen1986,Holland1993,Dowling1998,Kim1998,Pooser2004,Pezze2005,Pezze2006}%
. In all the above cases, the phase sensitivity $\Delta\phi$ has been
discussed in terms of two limits, known as the standard limit, $\Delta
\phi_{SL}=$ $1/\sqrt{N}$, and the Heisenberg limit\cite{Heisenberg1927},
$\Delta\phi_{HL}=$ $1/N$, where $N$ is the number of particles that enter the
interferometer during each measurement cycle.  These arguments are based on
results of standard estimation theory\cite{Helstrom1976} which connects an
ensemble of measurement outcomes, $m_{i}$, $i=1,2,\cdots,M$, with
corresponding phases, $\phi_{i}$, through a theoretical relation $m=m(\phi)$.
An example of the theoretical relation associated with some quantum observable
is 
$m(\phi )=\langle \phi |\widehat{m}|\phi \rangle $, where the state is parameterized by
a single parameter $\phi$. 
Standard estimation theory predicts that the standard deviation,
$\Delta\phi$, of the probability distribution for the phase $\phi$, is related
to the standard deviation in the measurements, $\Delta m$, by~\cite{Helstrom1976}
\begin{equation}
\Delta \phi=    \left| \frac{dm(\overline{\phi})}{d\phi} \right|^{-1} \, \Delta m.     \label{StandardDeviationsRelated}%
\end{equation}
Equation (\ref{StandardDeviationsRelated}) assumes that there is a single peak
in the phase probability density distribution $p(\phi)$, whose width can be
characterized by the standard deviation $\Delta\phi$. \ In general, a Bayesian
analysis of measurement outcomes $m$ for an interferometer can lead to a
conditional probability density distribution for the phase, $p(\phi|m)$, that
has multiple peaks.  Indeed, multiple peaks have been observed by Pezze and Smerzi~\cite{Pezze2005,Pezze2006} 
in the context 
of interferometry described by angular momentum algebra~\cite{YurkePRA1986,Kim1998}.
Therefore, the standard deviation $\Delta\phi$ is not an adequate metric to
characterize the phase sensitivity of an interferometer when multiple peaks are present in the 
phase probability distribution. 

In this Letter, we
propose to characterize the phase sensitivity of an interferometer by an alternative metric---the
fidelity---which is the Shannon mutual information~\cite{Shannon1948,NielsenChaung2000}, $H(\Phi \colon \! M)$,  
between the phase shift $\phi$\ and the measurement outcomes, $m$.
As an example, we consider the specific case of an optical
Mach-Zehnder interferometer in a sensor configuration, see Figure 1. We use an
exact Bayesian analysis to compute the conditional probability density
distribution for the phase $\phi$, $p(\phi|m)$, and we find that multiple peaks exist.
We compute the Shannon mutual information, $H(\Phi \colon \! M)$, for two types of input states, 
Fock states and {\it N00N}  states, which have been of great interest~\cite{Kapale2005,Mitchell2004}. 
We find that the fidelity associated with the Fock state input  is greater than for {\it N00N} state input.

{\it Phase Sensitivity.}
A quantum interferometer can act as a sensor of an external field $F$. 
A quantum state $|\Psi_{in}\rangle$ is
input into the interferometer and, through an interaction Hamiltonian
$H_{I}(F)$, the state interacts with an external classical field $F$, leading
to a phase-shifted output state $\ |\Psi(F)\rangle\,\ $that is
parameterized by the field $F$. \ We assume that a single parameter, the phase
shift\ $\phi$, is sufficient to describe the physics of the interaction
process. \ A general description of such a sensor can then be given in terms
of the scattering matrix, $S_{ij}(\phi)$, that connects the $N_p$ input-mode
field operators $\widehat{a}_{i}$ \ to the $N_p$  output-mode field operators
$\widehat{b}_{i}$,
\begin{equation}
\widehat{b}_{i}=\sum_{j=1}^{N_p}S_{ij}(\phi)\,\widehat{a}_{j} ,
\label{InputOutputDef}%
\end{equation}
where $i,j=1,2,\cdots,N_p$ and $\phi$ is the phase shift of the scattered
(output) state. \ The field $F$ leads to a phase shift $\phi$ of the scattered
state, whose detailed relation is determined by the interaction Hamiltonian
$H_{I}(F)$, which we do not consider any further here. \ The input state
evolves through the interferometer according to the unitary time-evolution
operator, \ $\widehat{U}(\phi)$, which relates the
input state at $t=-\,\infty\ $to the output state at $t=+\,\infty$,
\begin{equation}
|\Psi_{out}\rangle \,= \,\ \widehat{U}(\phi) \, |\Psi_{in}\rangle.  \label{UnitaryEvolution}%
\end{equation}

Measurements are described by a set of\ Positive Operator-Valued Measure (POVM)
operators, $\{\widehat{E}_{1},\widehat{E}_{2},\cdots,\widehat{E}_{N_m}\}$, where each operator \ $\widehat
{E}_{m}$ corresponds to a measurement outcome $m$. The conditional
probability of a given measurement outcome $m$ for a given phase shift $\phi$, is
the expectation value \ $P(m|\phi)=\langle \Psi_{in} |  \widehat
{E}_{m}(\phi) |\Psi_{in}\rangle$, where the Heisenberg operators
$\widehat{E}_{m}(\phi)$\ evolve in time and the states are constant.  
From Bayes' rule, we find the
conditional probability density, $p(\phi | m)$, for the phase shift $\phi$
for a given measurement outcome $m$ is 
\begin{equation}
p(\phi | m)=\frac{P(m | \phi)}{\int_{-\pi}^{+\pi}P(m | \phi^{\prime})\, \, d\phi^{\prime}},  \label{PhaseProbabiliyDensity}%
\end{equation}
where we have assumed a uniform {\it a priori} probability density for the phase shift $\phi
$. 

In order to have a good sensor,  the 
distribution $p(\phi|m)$ should have a narrow peak that is centered about some value of
the phase, for each measurement outcome $m$.  The phase sensitivity of an interferometer, or quantum interferometric
sensor, is usually taken to be the width of the single peak of the probability
density $p(\phi|m)$. 

A careful analysis of the probabilities $P(m|\phi)$ as functions of the scattering
matrix $S_{ij}(\phi)$ shows that in general the probabilities $P(m|\phi)$ are oscillatory
functions of $\phi$.  Consequently, the probability density for the phase,
$p(\phi|m)$, will have multiple peaks. The physics responsible for this
is due to the mutual symmetry  of the quantum state  and the
measuring apparatus (described by operators \ $\widehat{E}_{m}(\phi)$). 
\begin{figure}
\includegraphics{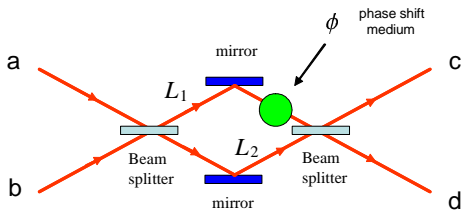}
\caption{A Mach-Zehnder interferometer is shown consisting of two 50-50 beam
splitters and two mirrors. The two input ports and two output ports are shown
along with a medium that induces the phase shift $\phi$. \ }%
\label{MachZehnderInterferometer}
\end{figure}
Since the probability density $p(\phi|m)$ has multiple peaks, the standard 
deviation $\Delta\phi$ is not an
adequate measure of the interferometer's phase sensitivity.  

We propose a new metric
for interferometer phase sensitivity--the fidelity--defined as the Shannon 
mutual information between the
set of possible phase values $\phi$, and the possible
measurement outcomes $m$.  For convenience, we discretize the phase shift
into values $\phi_{k}=\pi k/N_{\phi}$, for $k=\pm1,\pm2,\cdots\pm,N_{\phi}$,
and consider the mutual information between the $2N_{\phi}$ dimensional
alphabet of input phases, $\phi_{k}$, and the $N_m$-dimensional alphabet of output
measurement outcomes, $m=0,1,2,\cdots,N_m$. In the limit $N_{\phi}%
\rightarrow\infty$, the Shannon mutual information between the phase shift and
the measurements outcomes $m$ is given by%
\begin{widetext}
\begin{equation}
H(\Phi \colon \! M)=\frac{1}{2\pi}\sum_{m}\int_{-\pi}^{+\pi}d\,\phi\,\,P(m|\phi
)\,\,\log_{2}\left[  \frac{2\pi\,P(m|\phi)\,}{\int_{-\pi}^{+\pi}%
\,\,\,P(m|\phi^{\prime})\,\,d\,\phi^{\prime}}\right],
\label{MutualInformationEq}%
\end{equation}
\end{widetext}
where we have taken the {\it a priori} phase distribution $p(\phi)=1/(2 \pi)$ to be uniform over the 
interval \mbox{$-\pi < \phi \le \pi$}. 
The mutual information $H(\Phi \colon \! M)$ describes the amount of information, on average,
that an experimenter gains about the phase $\phi$  
on each use of the interferometer.  The mutual information depends on the type of input state and on
the type of measurement (POVM) performed.

{\it Mach-Zehnder Sensor.} As a specific example of the above discussion, 
we consider a generic optical
Mach-Zehnder interferometer, see Figure
\ref{MachZehnderInterferometer}.  The interferometer can be characterized by 
a scattering matrix%
\begin{eqnarray}
S_{ij}(\phi) & =  & \frac{1}{2}(e^{i\phi}e^{ikL_{1}} -e^{ikL_{2}})\sigma_{z} \nonumber\\
             &  & -\frac{i}{2}(e^{i\phi}e^{ikL_{1}}+e^{ikL_{2}})\sigma_{x} , \label{ScatteringMatrix}%
\end{eqnarray}
where%
\begin{equation}
\sigma_{x}=\left(
\begin{array}
[c]{cc}%
0 & 1\\
1 & 0
\end{array}
\right)  ,\sigma_{z}=\left(
\begin{array}
[c]{cc}%
1 & 0\\
0 & -1
\end{array}
\right) , \label{PaliMatrices}%
\end{equation}
 $L_{1}$ ($L_{2}$) is the upper (lower) path length through the
interferometer,  $k=\omega/c$,  $\omega$ is the angular frequency of
the photons and $c$ is the speed of light in vacuum. For any input state, the 
conditional probability for an outcome of observing $n_{c}$ and $n_{d}$
photons in output ports ``c" and ``d", respectively, for a given phase shift
$\phi$, is
\begin{equation}
P(n_{c},n_{d}\mid\phi)=\,\langle \Psi_{in}\,| \widehat{\pi
}_{n_{c},n_{d}}(\phi)  |\Psi_{in}\rangle ,
\end{equation}
where $\widehat{\pi}_{n_{c},n_{d}}(\phi) = |n_{c}, n_{d}\rangle \langle n_{c}, n_{d} |$
and $|n_{c}, n_{d} \rangle$ is the output state in the Schr\"odinger picture. 
For an $N$-photon Fock state input in port ``a",  
$ | \Psi_{in} \rangle=| N_{a}, 0_{b}\rangle $,
we find (taking $L_1 = L_2$)
\begin{eqnarray}
P_{N}(n_{c},n_{d}|\phi) = \phantom{xxxxx}  &   & \nonumber\\
                             \frac{N\,!}{n_{c}!\;n_{d}!} \,  \delta_{N,\,n_{c}\,+\,n_{d}} & & \! \! \! \! \! \!  \sin^{2n_{c}} (\frac{\phi}{2})  \cos^{2n_{d}%
}(\frac{\phi}{2}) . 
\end{eqnarray}
\begin{figure}
\includegraphics[scale=0.44,viewport= 0cm 0cm  20cm 12cm]{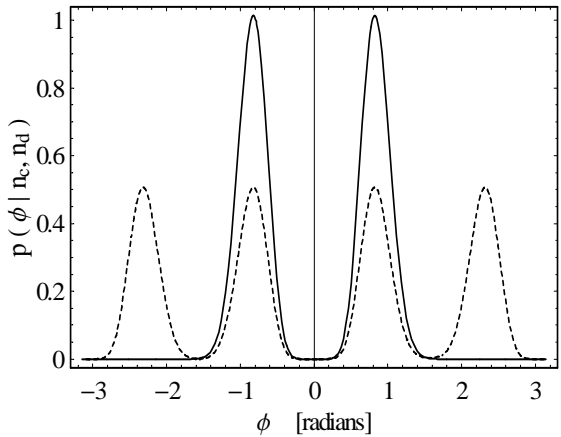}
\caption{Probability density of the phase $p(\phi | m)$ for Fock state input (solid curve) and $N00N$ state input (dashed curve) for $N=25$ photons
for measurement outcome $m =\{4_c, 21_d\}$.}%
\label{PhaseProbability0421}%
\end{figure}
Similarly, for a $N00N$-state input
\begin{equation}
|\Psi_{N00N}\rangle= \frac{1}{\sqrt{2}}\,\left[  \;|N_{a},0_{b}\rangle+|0_{a},N_{b}\rangle\;\right]
\end{equation}
the conditional probability is%
\begin{widetext}
\begin{equation}
P_{N00N}(n_{c},n_{d}|\phi)=\frac{1}{2\,}\frac{N\,!}{n_{c}!\;n_{d}!}%
\delta_{N,\,n_{c}\,+\,n_{d}}\,\left[  \sin^{n_{c}}(\frac{\phi}{2})\cos^{n_{d}%
}(\frac{\phi}{2})+(-1)^{n_{c}}\sin^{n_{d}}(\frac{\phi}{2})\cos^{n_{c}}%
(\frac{\phi}{2})\right]  ^{2} .
\end{equation}
\end{widetext}

It is clear that the probabilities 
$P_{N}(n_{c},n_{d}|\phi)$ and $P_{N00N}(n_{c},n_{d}|\phi)$, have
multiple peaks, and therefore the resulting probability densities for the phase,
$p(\phi| n_{c},n_{d})$ given by Eq.(\ref{PhaseProbabiliyDensity}), also have 
multiple peaks.  For a given $N$-photon Fock state input into port ``a" and vacuum
input into port ``b", the probability distribution
$p(\phi| n_{c},n_{d})$ has either one or
two peaks, depending on the measurement outcome. 
For the $N$-photon (entangled) $N00N$ state input,  the
probability distribution $p(\phi| n_{c},n_{d})$  has  one, two, three, or four peaks,
depending on the measurement outcome.  See Figure \ref{PhaseProbability0421} for an example  plot  of 
$p(\phi| n_{c},n_{d})$  for Fock state and $N00N$ state input for measurement outcome $\{4_c, 21_d\}$.
There is more
ambiguity in estimating the phase from the phase probability density for $N00N$ state
input than Fock state input, because there are more peaks.

For input states with increasing photon number, the probability densities,
$p(\phi| n_{c},n_{d})$, have narrower peaks, but the number of peaks remains the
same: one or two for Fock state input, and one, two, three, or four peaks for $N00N$
state input.

\begin{figure}
\includegraphics[scale=0.4,clip=true]{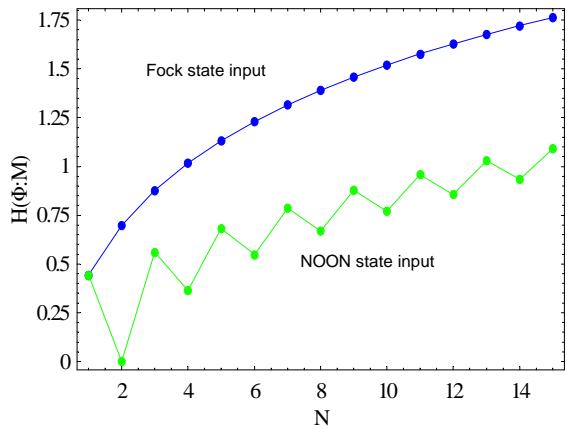}
\caption{The mutual information $H(\Phi \colon \! M)$ is plotted vs. photon number for
the case of N-photon Fock state input into port ``a" (blue dots) and for an N-photon $N00N$
state input (green dots). Lines connect successive photon number points.}%
\label{MutualInfoPlot}%
\end{figure}

When the interferometer is used as a sensor, it can be thought of as transmitting information about the phase
to the experimenter via each measurement outcome.  As described above,
due to multiple peaks in the phase distribution, 
 we do not attempt to use the width of the
probability distributions to describe the quality of this sensor. Instead, we use the Shannon mutual
information, given in Eq.(\ref{MutualInformationEq}), as a measure of the
fidelity of an interferometric sensor.  For the case of Fock state and
$N00N$\ state input, the mutual information, $H(\Phi \colon \! M)$, is plotted in Figure~\ref{MutualInfoPlot}.  
In both cases, the fidelity of the interferometer, acting like a
sensor, increases with increasing photon number due to the increased information carrying capacity of  a
higher-dimensional output alphabet
associated with the $N+1$ measurement outcomes $\{n_{c},n_{d}\}$. However, for
the same photon number input, the fidelity of the interferometer is clearly
greater for Fock state input than for $N00N$\ state input. This shows how the 
mutual information is sensitive to the number of peaks and not just the width $\Delta \phi$ of a single peak.
 Clearly, Fock
states can carry more information about the phase to the measurement outcomes
than $N00N$\ states. This striking result demonstrates that the use of entangled input states does not 
lead to improvement over Fock state input\cite{Kapale2005}.  

In order to optimize the Mach-Zehnder sensor, we can consider a more general class of states
$|\Psi_{in}\rangle =\sum\limits_{n=0}^{N}\,c_{n}\,|\,n_{a,}\,(N-n)_{b}\,\rangle $,
where $c_n$ are complex coefficients.  The sensor can be optimized by finding the $N+1$
coefficients that maximize the fidelity, $H(\Phi \colon \! M)$, subject to the normalization constraint
$\langle \Psi_{in}|\Psi_{in}\rangle =1$.

Work is in progress to analyze the mutual information for 
repeating the experiment $N$ times for arbitrary input states.
An interesting  example is the case when one-photon is input into port ``a"
and vacuum is input into port ``b". When the experiment is repeated $N$
times, with $M_{0\text{ }}$ outcomes $\{0_{c},1_{d}\}$ and $M_{1\text{ }}$
outcomes $\{1_{c},0_{d}\}$, where $N=$ $M_{0\text{ }}+M_{1\text{ }}$, 
the mutual information $H(\Phi \colon \! M)$ vs. $N$ is identical to that of
Fock state input with the same $N$.

{\it Summary.} We have considered a generic Mach-Zehnder optical interferometer operating as a 
sensor of a classical field.  Using a Bayesian analysis, we have shown that the 
conditional probability distribution for
the phase shift, $p(\phi| n_{c},n_{d})$, has multiple peaks and is not adequately
described by the standard deviation $\Delta\phi$, which has been used in discussion of the 
the standard limit ($\Delta\phi_{SL}\sim 1/\sqrt{N}$) and the Heisenberg limit ($\Delta\phi_{HL}\sim 1/N$) .  

We proposed an alternative metric--called the fidelity of the
interferometer--which is the Shannon mutual information, $H(\Phi \colon \! M)$, between
the phase shift  $\phi$ and the possible measurement outcomes $m$.  For an
interferometer used as a quantum sensor,  we have shown that the fidelity is a measure of
the quality of a sensor to detect external classical fields.  

For the case of a
generic Mach-Zehnder optical interferometer, we found the surprising result
that entangled $N00N$ state input leads to a lower fidelity than Fock state
input, for the same photon number. This result is intuitively understood 
because there are a larger number of peaks 
(bigger ambiguity in phase) in $p(\phi| n_{c},n_{d})$ for $N00N$ state input than for Fock state input.

The interferometer fidelity that we proposed is applicable to a wide 
variety of optical and matter wave interferometers,  
with arbitrary number of input/output ports.  
This measure of interferometer fidelity can be used as a metric for quantum interferometric 
sensors of classical fields, such as gravitational wave sensors, as well as optical gyroscopes and matter-wave
gyroscopes based on BEC.

\begin{acknowledgments}
This work was sponsored by the Disruptive Technology Office (DTO) and the Army Research Office (ARO).
This research was performed while P.\ L.\ held a National Research Council Research Associateship Award
at the U. S. Army Research Laboratory.  
\end{acknowledgments}

\end{document}